\documentstyle[12pt]{article}
\topmargin--0cm
\oddsidemargin--1mm
\begin{document}
\renewcommand{\thesubsection}{\arabic{subsection}}
\renewcommand{\theequation}{\thesubsection.\arabic{equation}}
\def\<{\langle}
\def\>{\rangle}
\newcommand{\pl}{\partial}
\def\R{\relax{\rm I\kern-.18em R}}
\def\1{\relax{\rm 1\kern-.27em I}}
\def\wne{\relax{\approx \kern-.93em /}\ }
\def\o{\overline}
\def\H{{\cal H}}
\def\D{{\cal D}}
\def\n{\nonumber}
\def\ra{\rightarrow}
\def\tint{{\textstyle\int}}
\newcommand{\mb}[1]{\mbox{\boldmath${\bf #1}$}}
\newcommand{\Z}{Z\!\!\! Z}
\newcommand{\ph}{PS_{ph}}
\newcommand{\oo}[1]{\stackrel{\circ}{\omega}\ \!\!\! ^{#1}}
\newcommand{\ooo}[1]{\stackrel{\circ}{\omega}\ \!\!\! _{#1}}

\begin{center}
{\LARGE Coordinate--Free Quantization of Second--Class

\vskip 0.1cm
Constraints}

\vskip 1cm
John R. KLAUDER\ $ ^a$\ \ and\ \
Sergei V. SHABANOV\ $ ^{b,}$\footnote{\noindent
Alexander von Humboldt fellow;
on leave from Laboratory of Theoretical
Physics, JINR, Dubna, Russia.}

\vskip 0.5cm
$ ^a${\it Departments of Physics and Mathematics, University of Florida,
Gainesville, 32611 USA

\vskip 0.2cm
$ ^b$Institute for Theoretical Physics, FU-Berlin,
Arnimallee 14, WE 2, D-14195, Berlin, Germany}.

\end{center}

\begin{abstract}
The conversion of second-class constraints into first-class
constraints is used to extend the coordinate-free path integral
quantization, achieved by a flat-space Brownian motion regularization
of the coherent-state path integral measure, 
to systems with second-class constraints
\end{abstract}

\subsection{Second-class constraints}
\setcounter{equation}0
On performing the Legendre transformation for generalized velocities
of a Lagrangian  of a dynamical system, one very often gets
relations between canonical coordinates and momenta that
do not involve time derivatives. They are therefore not equations
of motion and are called primary constraints \cite{dirac}. The primary
constraints should be satisfied as time proceeds, which leads to
further conditions on dynamical variables known as secondary constraints
\cite{dirac}.

Let $\varphi_a = \varphi_a(\theta)=0$ be all independent constraints
(primary and secondary) in the system; here $\theta^i,\
i=1,2,...,2N$, denote canonical
variables that span a Euclidean phase space of the system. The canonical
symplectic structure is assumed on the phase space
$\{\theta^i,\theta^j\} =\oo{ij}$; one can, for instance, set
$q^{n} = \theta^{2n-1}$ and $p_n = \theta^{2n},\ n=1,2, ... ,N$
for canonical coordinates and their momenta, then $\{p_n,q^{m}\}
=\delta_n^m$ and other components of the canonical symplectic structure
are zero. Let $H(\theta)$ be the canonical Hamiltonian of the system.
Since $\varphi_a$ is a complete set of constraints,
\begin{equation}
\dot{\varphi}_a = \{\varphi_a,H\} = C^b_a(\theta)\varphi_b \approx 0\ ,
\label{1.1}
\end{equation}
where the symbol $\approx$ implies the weak equality \cite{dirac} that
is valid on the constraint surface $\varphi_a =0$.

Systems with constraints admit a more general dynamical description
where the Hamiltonian can be replaced by a generalized one
$H_T=H+\lambda^a(\theta,t)\varphi _a(\theta)$ with $\lambda ^a$ being
arbitrary functions of $\theta ^i$ and time:
\begin{equation}
\dot{\theta }^i=\{\theta ^i,H_T\}\approx \{\theta ^i,H\}+\lambda ^a\{
\theta ^i,\varphi _a\}\ .
\label{1.2}
\end{equation}
The condition (\ref{1.1}) with $H$ replaced by $H_T$ enforces some
restrictions on $\lambda ^a$. Indeed
\begin{equation}
\dot{\varphi }_a\approx \{\varphi _a, \varphi _b \}\lambda ^b =0\ ,
\label{1.3}
\end{equation}
that is, the number of independent functions $\lambda ^a$ is determined by
the rank of the matrix $\{\varphi _a,\varphi _b \}= \Delta _{ab}$
on the surface of constraints.
If the rank is zero, i.e. $\{\varphi _a,\varphi _b \}=C_{ab}^c\varphi
_c\approx 0$ then all the $\lambda ^a$ are arbitrary, and the solutions
$\theta ^i(t)$ to the equations of motion (\ref{1.2}) would contain arbitrary
functions $\lambda ^a$. Such constraints are called first-class constraints
\cite{dirac} and transformations of $\theta ^i(t)$ generated by $\lambda ^a
\rightarrow \lambda ^a+\delta \lambda ^a$ are known as gauge transformations.
The latter implies in particular that the dynamical system has more
non-physical canonical variables in addition to those fixed by the
constraints $\varphi _a=0$. They can be removed by specifying $\lambda ^a$
(by gauge fixing).

If the rank of $\Delta _{ab}$ on the surface $\varphi_a=0$
is equal to the number of independent
constraints then $\det \Delta _{ab}\wne 0$ and all the $\lambda _a$ must
be zero. The system admits no gauge arbitrariness. Such constraints are
with an arbitrarily large diffusion constant
called second-class constraints \cite{dirac}. Of course there could
well be a ``mixed'' case when the system possesses second and first class
constraints (the matrix $\Delta _{ab}$ is degenerate, but $\Delta
_{ab}\wne 0$). In what follows only second-class constraints are
considered.

The number of second-class constraints must be even because the
determinant of an antisymmetric matrix of odd order is always zero. So we
set $a=1,2,\ldots, 2M$, that is the system has only $2(N-M)$ physical
canonical variables that describe dynamics on the physical phase space
determined by $\varphi _a(\theta )=0$.

\subsection{Local parametrizations of the physical phase space}
\setcounter{equation}0

One can introduce a local parametrization of the constraint surface
$\theta ^i=\theta
^i(\vartheta)$ where $\vartheta ^\alpha ,
\alpha =1,2,\ldots,2(N-M)$, are chosen so that 
$\varphi_a(\theta(\vartheta))$ identically vanish for all
values of $\vartheta^\alpha$. The local coordinates
$\vartheta^\alpha$
 span the
physical phase space. They may also serve
as local symplectic variables on the physical phase space, provided there
is an induced symplectic structure on it. 
To obtain an induced symplectic structure
for an ordinary change of variables
on phase space, one would have to invert
the relations $\theta ^i=\theta ^i(\vartheta )$ and calculate the Poisson
bracket of $\vartheta ^\alpha$. This procedure is incorrect for systems
with constraints. The inverse relations $\vartheta ^\alpha =\vartheta
^\alpha (\theta)$ are determined modulo the constraints $\varphi _a=0$.
Since $\{\theta ^i,\varphi _a\}\neq 0$, the Poisson bracket $\{\vartheta
^\alpha ,\vartheta ^\beta \}$ is ambiguous and 
does not induce the right symplectic
structure. Recall that due to the same reason ($\{\theta ^i,\varphi
^a\}\neq 0$), the constraints should not be solved before calculating the
Poisson bracket in equations of motion (\ref{1.2}).

The problem is resolved by means of the Dirac bracket 
that reads as \cite{dirac}
\begin{equation}
\{A,B\}_D=\{A,B\}-\{A,\varphi _a\}\Delta ^{ab}\{\varphi _b,B\}
\label{2.1}
\end{equation}
for any $A$ and $B$, and $\Delta _{ab}\Delta ^{bc}=\delta ^c_a$. The Dirac
bracket possesses three important properties
\cite{dirac}. First, it satisfies the
Jacobi identity, the Leibnitz rule and is antisymmetric, therefore it
determines a symplectic structure
\begin{equation}
\{\theta^i,\theta ^j\}_D=\ \oo{ij}-\{\theta ^i,\varphi _a\}\Delta ^{ab}\{
\varphi _b,\theta ^j\}\equiv \omega ^{ij}_D(\theta)\ .
\label{2.2}
\end{equation}
Second, it vanishes for any $A(\theta)$ and any of the constraints
$\varphi _a$,
\begin{equation}
\{A,\varphi _a\}_D=0\ .
\label{2.3}
\end{equation}
As a consequence of (\ref{2.3}) we deduce the third property
\begin{equation}
\dot{A}=\{\dot{A},H\}_D\approx \{A,H\}\ ,
\label{2.4}
\end{equation}
i.e. the equations of motion are not affected by the replacement of the
Poisson bracket by the Dirac one.

The advantage of using the Dirac bracket is that one can solve the
constraints at any stage of calculation, before or after calculating the
brackets. The latter is guaranteed by (\ref{2.3}). 
In particular, given a set of local parameters $\vartheta
^\alpha =\vartheta ^\alpha (\theta)$ spanning the surface $\varphi ^a=0$,
the induced symplectic structure is unique
\begin{equation}
\{\vartheta ^\alpha ,\vartheta ^\beta \}_D=\omega ^{\alpha
\beta}(\vartheta )\ .
\label{2.5}
\end{equation}
We remark that the Dirac symplectic
structure (\ref{2.2}) is degenerate. Its non-degenerate part is given by
(\ref{2.5}) in local coordinates $\vartheta ^\alpha$. There is an infinite
number of choices of local symplectic coordinates on the physical phase
space. One can find such a parametrization for which the induced
symplectic structure (\ref{2.5}) has the canonical form $\omega ^{\alpha
\beta}=\oo{\alpha \beta}$. The latter follows from the Darboux theorem,
and the corresponding $\vartheta ^\alpha$ are Darboux coordinates for the
symplectic structure (\ref{2.5}). But even the Darboux coordinates are not
unique, since they are determined up to a canonical transformation.

Thus, classical dynamics of second-class constrained systems exhibits a
physical-phase-space reparametrization invariance.

\subsection{Examples of second-class constraints}
\setcounter{equation}0

Consider a Lagrangian of the form
\begin{equation}
L=\frac{1}{2}\dot{\mb{x}}^2+yF(\mb{x})\ ,
\label{3.1}
\end{equation}
where $\mb{x}$ is a radius vector in $\R^N$.
It describes a motion of a point-like particle of unit mass on an
$N-1$- dimensional
surface determined by the equation 
$F(\mb{x})=0$, as follows from the Euler-Lagrange equation
for the Lagrange multiplier $y$. The canonical momentum for
$y$ vanishes yielding the primary constraint
\begin{equation}
p_y=\frac{\pl L}{\pl \dot{y}}=0\ .
\label{3.2}
\end{equation}
Doing the Legendre transformation for $\dot{\mb{x}}$, we arrive at the
Hamiltonian
\begin{equation}
H=\frac{1}{2}\mb{p}^2-yF(\mb{x})+\lambda p_y\ ,
\label{3.3}
\end{equation}
where $\mb{p}$ is the canonical momentum for $\mb{x}$, and $\lambda$ is an
arbitrary function of canonical variables and time. 
Its occurrence in (\ref{3.3}) is
due to (\ref{3.2}) (since $p_y$ vanishes identically, the canonical
Hamiltonian is determined up to any function that vanishes as $p_y=0$).

To find  secondary constraints, one should calculate the Poisson bracket
of the primary constraint (\ref{3.2}) with the Hamiltonian (\ref{3.3})
\begin{equation}
\dot{p}_y=\{p_y,H\} \approx F(\mb{x})=0\ ,
\label{3.4}
\end{equation}
that is, the motion is indeed constrained to the surface $F=0$. We should
continue checking the dynamical self-consistency for the constraint
(\ref{3.4})
\begin{equation}
\dot{F}(x)=\{F,H\}\approx (\mb{p},\mb{\pl}F)=0\ .
\label{3.5}
\end{equation}
Equation (\ref{3.5}) determines a new constraint. Let us denote the
constraints (\ref{3.2}), (\ref{3.4}) and (\ref{3.5}) as $\varphi
_{1,2,3}$, respectively. Then there must be
\begin{equation}
\dot{\varphi}_3=\{\varphi _3,H\}\approx p_ip_j\pl _i\pl
_jF+y(\mb{\pl}F)^2 =0
\label{3.6}
\end{equation}
The new constraint (\ref{3.6}) is denoted by $\varphi _4$. The theory does
not have more constraints because the condition $\dot{\varphi}_4=\{\varphi
_4,H\}\approx 0$ 
yields an equation for an arbitrary function $\lambda$, rather
than for canonical variables. It is not hard to be convinced that all four
independent constraints $\varphi _a$ are of the second class, i.e. $\det
\Delta _{ab}=\det \{\varphi _a,\varphi _b\}\wne 0$.

A geometrical meaning of (\ref{3.4}) and (\ref{3.5}) is transparent.
Equation (\ref{3.4}) implies that the particle moves along the surface
$F=0$. Equation (\ref{3.5}) means that the particle momentum remains 
tangent to
the surface $F=0$ during the motion. 
Note that the vector $\mb{\pl}F$ is locally transverse
to the surface $F=0$. The constraints (\ref{3.2}) and (\ref{3.6}) are
artifacts of constructing the Hamiltonian formalism from the Lagrangian
(\ref{3.1}) where $y$ is not a dynamical variable, rather it is a Lagrange
multiplier used to enforce the constraint (\ref{3.4}) in the Lagrangian
formalism. Since in the Legendre transformation the variable $y$ is
treated as an independent dynamical variable, the associated Hamiltonian
formalism exhibit two extra constraints (\ref{3.2}) and (\ref{3.6}) to
suppress dynamics of $y$ and $p_y$.

In fact, we may start right from the Hamiltonian formalism
(\ref{3.3}), setting in it $\lambda \equiv 0$ and $y$ to be an arbitrary
function of $\mb{p}$ and $\mb{x}$. Then the constraint (\ref{3.4}) should
be regarded as the "primary" constraint. In this simplified approach
equation (\ref{3.6}) is not a constraint, but the equation for an arbitrary
function $y$ (the Lagrange multiplier). The Dirac bracket formalism leads
to the same answer for the symplectic structure on the physical phase
space, so we prefer the simplified Hamiltonian formalism. In general,
given a set of the second class constraints $\varphi _a$ to be imposed on
the motion generated by the Hamiltonian $H_s$ of a system under
consideration, one can consider a generalized Hamiltonian formalism
\begin{equation}
H=H_s(\theta)+\lambda ^a(\theta,t) \varphi _a(\theta )\ .
\label{3.7}
\end{equation}
The consistency conditions $\dot{\varphi}_a=\{\varphi _a,H\}
\approx 0$ yield
equations for arbitrary functions $\lambda _a$ whose solutions are
\begin{equation}
\lambda ^a =\Delta ^{ab}\{\varphi _b,H_s\}\ .
\label{3.8}
\end{equation}
For instance, to describe the motion along a surface $F=0$ in the
Hamiltonian formalism, one can take $H_s=\mb{p}^2/2$ and two constraints
$\varphi _1=F$ and $\varphi _2=(\mb{p},\mb{\pl}F)$ and consider
the generalized Hamiltonian dynamics (\ref{3.7}).

A symplectic structure on the physical phase space determined
by the constraints $\varphi_{1,2} =0$ is induced by the Dirac 
bracket
\begin{eqnarray}
\{x_i,x_j\}_D &=&0\ ;\label{xx}\\
\{x_i,p_j\}_D &=&\delta_{ij} - n_in_j\ ;\label{xp}\\
\{p_i,p_j\}_D &=&p_k\left(
n_j\pl_kn_i - n_i\pl_kn_j
\right)\ ,\label{pp}
\end{eqnarray}
where $n_i = n_i(\mb{x}) = \pl_iF/|\pl F|$ is a unit vector that
coincides with the normal to the surface when $\mb{x}$ is on
the surface. The symplectic structure is by construction degenerate.
To construct the induced (non-degenerate) symplectic 
structure on the physical phase space, one should introduce a local
parametrization of the constraint surface $\varphi_{1,2}=0$
and calculate the Dirac bracket for local coordinates spanning
the constraint surface.
For instance, given a local parametrization of the surface $F(\mb{x})
=0$ in the form $\mb{x} = \mb{x}^F(u),\ u\in \R^{N-1}$, the
physical momenta are $p_i = e_i^a(u)p_{u_a}$ where the vectors
$\mb{e}^a(u)$ form a basis in the tangent space of the surface.
In particular, one can take $e_i^a(u)=\pl x^F_i/\pl u_a$. From
the identity $F(\mb{x}^F)\equiv 0$ follows the orthogonality
relation $(\mb{n}(\mb{x}^F),\mb{e}^a)=0$. The variables $p_{u_a},
u_a$ serve as local coordinates on the physical phase space. The
corresponding symplectic structure is induced by
(\ref{xx})--(\ref{pp}).
Another possibility would be to solve $F=0$ with respect to, say,
$x_N$ and $\varphi_2 =0$ with respect to any of the momenta, say,
$p_N$, i.e. $u_a= x_a, p_{u_a}=p_a,\ a=1,2,...,N-1$.

Let us illustrate the procedure with the two-dimensional case,
the motion on a plane constrained to a curve.
We introduce the following 
parametrization of the constraint surface
\begin{equation}
\mb{x}=\mb{f}(u)\ ,\ \ \mb{p}=p_u\pl _u\mb{f}\ ,
\label{3.9}
\end{equation}
where $u$ is a parameter on the curve, note that $\mb{\pl}F\sim T\pl
_u\mb{f}$ along the curve $F=0$, where $T_{ij}=\varepsilon
_{ij}=-\varepsilon _{ji},\ \varepsilon _{12}=1$. The variables $u$ and
$p_u$ are local coordinates on the physical phase space. We get
\begin{eqnarray}
\Delta _{ab}&= & \{\varphi _a,\varphi _b\}=T_{ab}(\mb{\pl}F)^2\;,
\label{3.10} \\
\Delta ^{ab}&= & -T_{ab}(\mb{\pl}F)^{-2}\ .
\label{3.11}
\end{eqnarray}
Choosing some function $u=u(\mb{x})$ (e.g. one can simply invert the
relation $x_1=f_1(u)$) we obtain
\begin{equation}
p_u=\gamma (\mb{x}) (\mb{p},T\mb{\pl}F)\;,
\label{3.12}
\end{equation}
where $\gamma (\mb{x})$ depends on the choice of $u(\mb{x})$. Hence
\begin{equation}
\{u,p_u\}_D=\gamma (\mb{x})(\mb{\pl}u(x),T\mb{\pl}F)\vert _{F=0}=\Gamma
(u)\ .
\label{3.13}
\end{equation}
The Darboux transformation for the physical symplectic form reads
$(u,p_u)\rightarrow (u,p_u/\Gamma)$.

In general, the constraint surface may have a nontrivial topology,
which must be taken into account when studying the dynamics in
local symplectic coordinates.
Consider, for example, a particle on a circle, $F(\mb{x})=\mb{x}^2-R^2$.
The radial motion is frozen and the other constraint is
$(\mb{p},\mb{x})=0$. A natural parametrization of the physical phase space
(according to (\ref{3.9})) is
\begin{equation}
\mb{x}=Re^{\omega T}\mb{\chi}\ ,\ \ \mb{p}=p_\omega T\mb{x}(\omega )/R
\label{3.14}
\end{equation}
where $\chi _i=\delta _{i1}$ and $\omega =\tan ^{-1}x_2/x_1$, $p_\omega
=(\mb{p},T\mb{x})$. Calculating the Dirac bracket we obtain
\begin{equation}
\{\omega ,p_\omega \}_D =1\ ,
\label{3.15}
\end{equation}
i.e. $\omega $ and $p_\omega $ are Darboux coordinates on the physical
phase space which has the topology of a cylinder because $\omega$ is a
cyclic (compact) variable, $\omega \in [-\pi ,\pi )$. One can also choose
$u=x_1$ and $p_u=p_1$ as local coordinates 
on the physical phase space . The
corresponding symplectic structure assumes the form
\begin{equation}
\{u,p_u\}_D=1-\frac{u^2}{R^2}\ .
\label{3.16}
\end{equation}
The Darboux transformation for (\ref{3.16}) is
\begin{equation}
u=R\cos \omega\ ,\ \ p_u=-\frac{1}{R}p_\omega \sin \omega\ .
\label{3.17}
\end{equation}
Now the topology of the physical phase space is hidden in singularities of
the symplectic structure: It vanishes at $u=\pm R$.

The examples illustrates an arbitrariness in choosing a
parametrization of the physical phase space: The parametrization
is determined up to a general coordinate transformation
on the physical phase space. The Dirac bracket
ensures a covariance of the Hamiltonian dynamics with respect
to such transformations. This covariance of the classical dynamics
is lost upon canonical quantization as we now proceed to demonstrate.

\subsection{Ambiguities in the canonical quantization of $\ $second-class
constraints}
\setcounter{equation}0

The canonical quantization of a classical system implies that the
canonical variables $\theta ^i$ become hermitian operators $\hat{\theta
}^i$ that act in a Hilbert space and obey the canonical commutation
relations $[\hat{\theta }^i, \hat{\theta}^j]=i\hbar
\{\hat{\theta}^i,\hat{\theta}^j\}=i\hbar \oo{ij}$. The recipe is
generally correct only in Cartesian coordinates \cite{dirac2}. Though the
canonical variables $\theta ^j$ have been assumed to be Cartesian, the
quantization postulate should be modified when second-class constraints
are present. The point is that the conditions $\hat{\varphi}_a=\varphi
_a(\hat{\theta})=0$ can not be imposed on the operator level because they
would be in conflict with the commutation relations $[\hat{\varphi
}_a,\hat{\varphi }_b]\neq 0$: The constraints cannot be solved before
calculating the commutation relations. The problem is resolved by
replacing the Poisson bracket by the Dirac one in the canonical
quantization postulate, that is \cite{dirac}
\begin{equation}
[\hat{\theta}^i,\hat{\theta}^j]=i\hbar \{\theta ^i,\theta ^j\}_D\vert
_{\theta =\hat{\theta}}=i\hbar \omega ^{ij}_D (\hat{\theta })\ .
\label{4.1}
\end{equation}
Since the Dirac symplectic structure is degenerate, not every canonical
variable can be made an operator (e.g. if $\varphi _1=q=0,\
\varphi _2=p=0$, then the Dirac rule leads to $[\hat{q},\hat{p}]=0$, i.e.
the canonical variables are commuatative and, therefore, remain
$c$-numbers upon quantization).
For a generic second-class constrained system, the Dirac 
commutation relations (\ref{4.1}) are constructed so that
the operators of constraints commute with canonical variables
$[\hat{\varphi}_a,\hat{\theta}^i]=0$ and, hence, can be given
any $c$-number value, in paticular, $\hat{\varphi}_a =0$,
enforcing the constraints on the quantum level. This comprises
the geometrical meaning of the Dirac approach.

The recipe (\ref{4.1}) is not however free of ambiguities either. The
Dirac symplectic structure depends on the canonical variables and
therefore the replacement $\theta ^i$ by the corresponding operators
usually leads to the operator ordering ambiguity.

An incorrect operator ordering in the right-hand side of (\ref{4.1}) can
break the associativity of the operator algebra (\ref{4.1}) (the Jacobi
identity is violated upon quantization). To restore the associativity,
terms of higher orders of $\hbar$ should be added to the right-hand side
of (\ref{4.1}). Even after the associativity problem has been resolved in
some way, one needs still to verify that this solution has not violated
the quantization consistency conditions
$[\hat{\theta}^i,\hat{\varphi}_a]=0$. The latter would involve solving the
operator ordering problem in the constraints in a way compatible with the
operator ordering in the symplectic structure. In general this program
may be very involved. After all the consistency problems have been
resolved, one should face a not less difficult problem of constructing a
representation of the algebra (\ref{4.1}) in order to be able to calculate
amplitudes (e.g. the evolution operator kernel).

As an alternative approach one can consider quantization after solving the
constraints, meaning that the physical symplectic structure (\ref{2.5}) is
to be quantized. It should be noted that such an approach, though it resolves
the operator ordering in the constraints, still has this problem in the
symplectic structure (\ref{2.5}). Going over to the Darboux variables does
not help in this regard because  canonical quantization is not
generally correct in curvilinear coordinates as has been mentioned before. The
problem appears even more serious if one notices that the Darboux
coordinates are determined up to a canonical transformation, whereas 
canonical quantization and canonical transformations are not commutative
operations. Thus, such a reduced phase-space quantization is
coordinate dependent, and in this regard cannot be considered as
a self-consistent quantization scheme.

As an illustration, consider the canonical quantization of the Dirac
symplectic structure for a particle on a circle.
Here $\mb{n} =\mb{x}/|\mb{x}|$, and canonical quantization of 
the Dirac bracket
(\ref{xx}) -- (\ref{pp}) yields the following commutation relations 
\begin{eqnarray}
[\hat{x}_i,\hat{x}_j] &=& 0\ ,
\label{4.2} \\
\left[ \hat{x}_j,\hat{p}_k \right] &=& i\hbar (\delta _{jk}-\frac{\hat{x}_j
\hat{x}_k}{ \hat{\mb{x}}^2})\ ,
\label{4.3} \\
\left[ \hat{p}_j,\hat{p}_k \right] &=& i\hbar \frac{1}{\hat{\mb{x}}^2}
(\hat{p}_j \hat{x}_k-\hat{p}_k \hat{x}_j)\ .
\label{4.4}
\end{eqnarray}
The operator ordering problem appears only in (\ref{4.4}) and can be
resolved by putting all the $\hat{p}_i$ either to the left or to the right
of the $\hat{x}_i$ in the right-hand side of (\ref{4.4}). Note that
$\hat{\mb{x}}^2$ commutes with all the canonical operators, so it does not
matter where it is placed in the right-hand side of (\ref{4.4}). One can check
that this ordering is compatible with the hermiticity of the canonical
operators and provides the Jacobi identity (associativity) in quantum
theory. In this particular model, the operator ordering problem in the
constraints is not of any relevance. Indeed,
$\hat{\varphi}_1=\hat{\mb{x}}^2-R^2$ does not have any, and commutes
with all the canonical operators. According to (\ref{4.3}), the ordering
correction to $\hat{\varphi}_2$ is a $c$-number and, hence, does not
affect the relations
$[\hat{\varphi}_2,\hat{x}_j]=[\hat{\varphi}_2,\hat{p}_j]=0$.

The algebra (\ref{4.2})--(\ref{4.4}) has a representation in a space of
$2\pi$-periodic functions $\psi(\omega +2\pi )=\psi (\omega )$
\begin{eqnarray}
\hat{x}_1\psi(\omega)&= & R\cos \omega \psi (\omega)\ ,
\  \ \ \ \ \hat{x}_2\psi (
\omega)=R\sin \omega \psi (\omega)\ ;
\label{4.5} \\
\hat{p}_1\psi (\omega)&= & \frac{i\hbar}{2R}(\sin \omega \pl _\omega +
\pl _\omega \sin \omega )\psi (\omega)\ ;
\label{4.6}\\
\hat{p}_2\psi (\omega)&=& -\frac{i\hbar}{2R}(\cos \omega \pl _\omega +
\pl _\omega \cos \omega )\psi (\omega)\ ;
\label{4.7}\\
\langle \psi _1|\psi _2\rangle &=&\int_{0}^{2\pi}d\omega \,\psi
^*_1\psi _2\ .
\label{4.8}
\end{eqnarray}
In this representation
$\hat{\varphi}_1=0$ and $\hat{\varphi}_2$ is a $c$-number determined by
the chosen operator ordering. The physical Hamiltonian assumes the form
\begin{equation}
\hat{H}_{ph}=\frac{1}{2R^2}\hat{p}^2_\omega+\frac{\hbar ^2}{8R^2}\ ,\ \
\hat{p}_\omega =-i\hbar \pl _\omega\ .
\label{4.9}
\end{equation}
In addition to the kinetic energy opertor on the circle, the
physical Hamiltonian contains a ``quantum'' potential that
has occured through the Dirac degenerate commutation relations.
For a generic manifold, the Dirac approach leads to a 
quantum potential that depends on position on the manifold.

A similar quantum potential was also predicted in the framework
of the path integral qunatization on manifolds \cite{dewitt},
and found to be proportional to the scalar curvature of the
manifold. 
It is interesting to observe that canonical quantization of the
Dirac bracket leads to a {\em different} prediction.
For an $N$-dimensional sphere, the scalar curvature potential
reads 
$\alpha\hbar^2 N(N-1)/R^2$ with $\alpha $ being a constant, i.e.,
it vanishes for a circle, whereas
the embedding of the $N$-sphere into $\R^{N+1}$ and canonical
quantization
of the Dirac bracket (\ref{4.2}) -- (\ref{4.4}) 
would yield the other form of the 
vacuum energy $\hbar^2 N^2/8R^2$. Note that the algebra 
(\ref{4.2}) -- (\ref{4.4}) applies
to quantum motion on the $N$-sphere as has been shown above.
Its representation is easy to find by going over to the spherical
coordinates and thereby obtaining the physical Hamiltonian.

Let us turn to quantization in Darboux variables.
The canonical quantization of the physical symplectic structure
(\ref{2.5}) in the Darboux coordinates would lead to a different
Hamiltonian
\begin{equation}
\hat{H}'_{ph}=\frac{1}{2R^2}\hat{p}^2_\omega\ .
\label{4.10}
\end{equation}
In this case the ``extra'' quantum potential is not unique at all. 
The source of troubles is that the Darboux variables (or any set of
canonical coordinates parametrizing the physical phase space) are
determined only up to a general canonical transformation and, 
hence, are generally
associated with non-Cartesian coordinates for which the canonical
quantization is not generally consistent \cite{dirac2}. For instance, one can
choose an alternative parametrization of the physical phase space
by going over to new canonical coordinates
$( \omega ,p_\omega )\rightarrow
(u=\sin \omega,\ p_u=p_\omega /\cos \omega ),\
\{u,p_u\} =1 $. After canonical
quantization, $[\hat{u},\hat{p}_u] = i\hbar$, 
the physical Hamiltonian would exhibit an operator ordering
problem which has no unique solution. 
Given a particular operator ordering in the Hamiltonian,
the change of variables
$u=\sin \omega$ would not generally lead to (\ref{4.10}),
rather the quantum Hamiltonian will have an $\omega$ dependent
potential of order $\hbar^2$. This quantum potential
is fully determined by the physical phase space parametrization
chosen. In this regard 
quantization of the Dirac degenerate symplectic structure (\ref{2.2}) in a
flat phase space looks more preferable because different parametrizations
of the physical phase space are associated with different realizations of
the same algebra of commutation relations for the canonical variables.

We would also like to mention that we do not consider the physical
origin of the second-class constraints and assume the latter
to be given a priori. In general, the applicability of the
Dirac formalism to concrete physical sysytems 
can be questioned. For instance, the motion on
a sphere can be physically interpreted as 
a motion in a thin spherical
layer so that the radial motion is confined by a spherically
symmetrical potential well.
By going over to spherical coordinates in the free-particle
Schr\"odinder equation, one can show that in the
limit when the layer thickness is much less than the
sphere radius, the quantum potential
is given by the quantum centrifugal barrier
$\hbar^2 N(N-2)/8R^2$ acting on a particle in the thin spherical
layer, which is different from the one predicted by the Dirac
or path integral approaches. 

So, it is an open question whether or not the Dirac quantization
scheme for second-class constraints can be applied to a particular
dynamical system.

\subsection{Quantization of second class constraints via 
the abelian conversion method}
\setcounter{equation}0

We have seen that the quantization of the Dirac bracket poses 
a few problems:
the operator ordering in the commutation relation algebra
(the associativity problem), the problem of finding
a representation of the Dirac commutation relations 
and the ordering
problem in the operators of constraints. Quantum dynamics depends on a
particular solution to them and, generally, is not unique. The first
two problems are most difficult. 
So one should develop a formalism that would
allow one to avoid them. Such a formalism is known as the conversion of
second-class constraints into first-class constraints by extending the
original phase space by extra (gauge) degrees of freedom \cite{conv}.

Let us extend the original phase space of a system with $2M$ independent
second-class constraints by adding to it $2M$ independent variables
$\phi^a$ with the canonical symplectic structure
\begin{equation}
\{\phi ^a,\phi ^b\}=\oo{ab},\ \ \{\phi ^a,\theta ^i\}=0\ .
\label{5.1}
\end{equation}
The original second-class constraints $\varphi _a(\theta)$ are then
converted into abelian first-class constraints $\sigma _a=\sigma _a(\theta
,\phi)$. Their explicit form is determined by a system of first-order
differential equations
\begin{equation}
\{\sigma _a,\sigma _b\}=0
\label{5.2}
\end{equation}
with the initial condition
\begin{equation}
\sigma _a(\theta ,\phi =0)=\varphi _a(\theta)\ .
\label{5.3}
\end{equation}
A dynamical equivalence of the original second-class constrained system to
the abelian gauge system is achieved by a specific extension of the
original Hamiltonian of the system,
$H_s(\theta)\rightarrow\bar{H}_s(\theta ,\phi)$, such that
\begin{equation}
\{\bar{H}_s, \sigma _a\}=0\ ,\ \ \bar{H}_s(\theta ,\phi =0)=H_s(\theta
)\ .
\label{5.4}
\end{equation}
One can show that equations of motion generated by the extended action
\begin{equation}
\bar{S}=\int\limits_{}^{}dt\left(\frac{1}{2}\phi ^a\ooo{ab}\dot{\phi }_b
+\frac{1}{2}\theta ^i\ooo{ij}\dot{\theta }^j-\bar{H}_s-\bar{\lambda }^a
\sigma _a\right)\ ,
\label{5.5}
\end{equation}
are equivalent to those generated by the original action
\begin{equation}
S=\int\limits_{}^{}dt\left(\frac{1}{2}\theta ^i\ooo{ij}\dot{\theta}^j
- -H_s-\lambda ^a\varphi _a\right)\ .
\label{5.6}
\end{equation}
We remark that in contrast to the theory (\ref{5.6}), the Lagrange
multipliers $\bar{\lambda}^a$ in the gauge theory (\ref{5.5}) are not
determined by the  equations of motion ( the matrix $\Delta _{ab}=\{\sigma
_a,\sigma _b\}$ vanishes according to (\ref{5.2})). A choice of
$\bar{\lambda }^a$ implies gauge fixing. In particular, one can always
choose $\bar{\lambda}^a$ so that $\phi ^a=0$ on
the constraint surface $\sigma_a =0$ for all moments of time. With
this choice the equations of motion of the system (\ref{5.5}) become the
equations of motion of the original system.

Equations (\ref{5.2}) and (\ref{5.4}) are not easy to solve for generic
$\varphi _a$ and $H_s$. However, if at least one set of Darboux variables
for the Dirac bracket is known, then the solution can be found explicitly
\cite{book}. In particular, for a particle on a circle we find
\begin{eqnarray}
\sigma _1& =& \varphi _1+P=\mb{x}^2-R^2+P \ ,\label{5.7} \\
\sigma _2&= & \varphi _2+2\mb{x}^2Q =(\mb{x},\mb{p})+2\mb{x}^2Q\ ,
\label{5.8}
\end{eqnarray}
where $\phi ^1=Q,\ \phi ^2=P$ and $\{Q,P\}=1$. To find a solution to
(\ref{5.4}), one can make use of a simple observation that $\sigma _{1,2}$
and $p_\omega =(\mb{p},T\mb{x})$ could be regarded as canonical momenta
for $Q$, $\ln r/R$ and $\omega =\tan ^{-1}x_2/x_1$, respectively. In
these new canonical variables equation (\ref{5.4}) is greatly simplified
and we obtain
\begin{equation}
\bar{H}_s=\frac{1}{2}\left(\frac{\sigma _2^2}{\sigma _1+R^2}+
\frac{(\mb{p},T\mb{x})^2}{\sigma _1+R^2}\right)\ .
\label{5.9}
\end{equation}
When $P=Q=0$, $\sigma _1+R^2=\mb{x}^2$ and $\sigma _2=(\mb{p},\mb{x})$ so
that $\bar{H}_s$ turns into the free-particle Hamiltonian
$H_s=\mb{p}^2/2$ (recall that the vectors $\mb{x}$ and $T\mb{x}$ form an
orthogonal basis on a plane). We remark also that formulas 
(\ref{5.7})--(\ref{5.9}) apply to an $N$-dimensional rotator (a motion
on the $N$-dimensional sphere); one should only replace the squared
total angular momentum of the rotator by its $N$-dimensional analog:
$(\mb{p},T\mb{x})^2 \rightarrow L^2 = \sum_a (\mb{p},T_a\mb{x})^2$
where $T_a$ are $N\times N$ real antisymmetric matrices, generators
of SO(N).

The first-class constraints (\ref{5.7}) and (\ref{5.8}) generate
gauge transformations on the extended phase space which are
translations of $Q$ and dilatation of the radial variables $r$, while
the angular variables and its momenta remain invariant. Thus,
$Q$ and $r$ are pure gauge degrees of freedom, and the angular
variables comprise guage invariant physical degrees of freedom
as expected. 

The gauge transformations are also canonical transformations with
generators being $\sigma _{1,2}$. An operator of finite canonical
transformations generated  by the constraints (\ref{5.7}) and (\ref{5.8})
has the form
\begin{equation}
\exp (\xi ^a {\rm ad}\sigma _a)\ ,\ \ 
{\rm ad}\sigma _a=\{\sigma _a,\cdot\}\ .
\label{5.10}
\end{equation}
Applying it to the canonical variables in the extended phase space, one
finds
\begin{eqnarray}
Q&\rightarrow &Q-\xi_1=Q_\xi\ ,\ \ \ \ P\rightarrow P+(1-e^{-2\xi
_2})\mb{x}^2 =P_\xi\ ;
\label{5.11} \\
\mb{x}& \rightarrow& e^{-\xi
_2}\mb{x}=\mb{x}_\xi\ ,\ \ \ \hskip.5cm \mb{p}\rightarrow e^{\xi
_2}\mb{p}+2\mb{x}(Qe^{\xi _2} - (Q-\xi _1)e^{-\xi _2})=\mb{p}_\xi \ .
\label{5.12}
\end{eqnarray}
{}From (\ref{5.11}) follows that there always exists a choice of the gauge
parameters $\xi _a$ such that $Q=P=0$ for all moments of time.
 
After the conversion has been made, the system can be quantized according
to the Dirac method for the gauge theories. Namely, all the canonical
variables of the extended Euclidean 
phase space become operators obeying the
standard Heisenberg commutation relations
\begin{equation}
[\hat{\theta}^j,\hat{\theta}^k]=i\hbar \oo{jk}\ ,\ [\hat{\phi}^a,
\hat{\phi}^b]=i\hbar\oo{ab}\ ,
\label{5.13}
\end{equation}
while the operators of constraints $\hat{\sigma }_a$ select physical
states
\begin{equation}
\hat{\sigma}_a\Psi_{ph}=0\ .
\label{5.14}
\end{equation}
Equation (\ref{5.14}) means that the physical states must be 
invariant under gauge transformations generated by first-class
constraints.
In particular, for the rotator we find that solutions to the
Dirac constraint equations (\ref{5.14}) are given by functions
$\Psi_{ph}=f(Q,r)\psi (\omega )$ where $f(Q,r)$ is uniquely fixed by
(\ref{5.14}), while $\psi (\omega)$ is an arbitrary function of the polar
angle on a plane. For the $N$-dimensional sphere,
the physical Hilbert space consists of functions on the sphere. 
So, it is the Hilbert space of a
quantum rotator as expected.

Thus, the problem of quantization of second class constraints 
has a natural gemetrical solution in the framework of the
conversion method. The technical difficulties do not disappear
completely; they are now associated with solving the conversion
equations (\ref{5.2}) and (\ref{5.4}). Nevertheless, the approach 
may be simpler 
than the original Dirac approach. Even in the case of the
rotator, the representation problem for the algebra 
(\ref{4.2})--(\ref{4.4}) does not appear to be a feasible task if
the geometrical origin of this algebra is unknown. 

An important advantage of the conversion method is that it does not rely
on any particular parametrization of the physical phase space.
For this reason we shall adopt it as starting point to develop
a path integral formalism for second-class constrained systems
which is invariant with respect to the parametrization choice of
the physical phase space and, in this sense, to achieve coordinate
independence of the quantum theory.

\subsection{The projection method}
\setcounter{equation}0

We assume the operators $\hat{\sigma}_a$ to be hermitian and that they
generate unitary transformations in the total Hilbert space. Since
by construction they commute with the Hamiltonian,
the total Hilbert space of an abelian gauge system obtained by the
conversion procedure can always be split into an orthogonal sum of a
subspace formed by gauge invariant states (\ref{5.14}) and a subspace that
consists of gauge variant states. 
Therefore an averaging of any state, being a linear
combination of eigenstates of the Hamiltonian, over the abelian
gauge group automatically leads to a projection operator onto the physical
subspace of gauge invariant states:
\begin{equation}
\hat{\cal P} = \int \delta_\sigma \Omega e^{-i\Omega^a\hat{\sigma}_a}
\label{projector}
\end{equation}
where $\delta_\sigma\Omega$ is a normalized measure on the space of gauge
transformation parameters. If the spectrum of the constraint operators
$\hat{\sigma}_a$ is not discrete, then the parameters $\Omega^a$ range over a
non-compact domain. In this case we adopt a certain regularization of the
measure $\delta_\sigma\Omega$ which provides \cite{kl1} (see also
Section 9)
\begin{equation}
\int\limits_{}^{}\delta_\sigma \Omega = 1
\label{6.3}
\end{equation}
and, hence, $\hat{\cal P}$ is the projection operator $\hat{\cal
P}^2=\hat{\cal P}$ such that
\begin{equation}
\hat{\cal P}\Psi _{ph}=\Psi _{ph}\ ,\ \ \ \ 
\hat{\cal P}\Psi_{nph} = 0
\label{6.4}
\end{equation}
for any gauge invariant state $\Phi_{ph}$ and any gauge variant
state $\Psi_{nph}$ (by definition, $\hat{\sigma}_a\Psi_{nph}
\neq 0$). Its kernel is determined as the gauge group
average of the unit operator kernel
\begin{equation}
\<\theta '',\phi ''|\theta ',\phi '\>^{ph}\equiv \<\theta '',\phi
''|\hat{\cal P}|\theta ',\phi '\>= \int\limits_{}^{}
\delta_\sigma\Omega\<\theta '',
\phi ''|e^{-i\Omega ^a\hat{\sigma }_a}|\theta ',\phi '\>\ ,
\label{6.5}
\end{equation}
where $|\theta,\phi\>$ is the coherent state defined as
\begin{equation}
|\theta,\phi\> = \exp\left(
i\theta^j\ooo{jk}\hat{\theta}^k + i\phi^a\ooo{ab}\hat{\phi}^b
\right)|0\>
\label{coh}
\end{equation}
with $|0\>$ being the ground state of the harmonic oscillator.

Accordingly, the physical transition amplitude
in the coherent-state representation is
obtained from the unconstrained one by averaging the latter over the gauge
group
\begin{eqnarray}
\<\theta '',\phi '',T|\theta ',\phi '\>^{ph}&= & \int\limits_{}^{}
\delta_\sigma\Omega\<\theta '',\phi '',T|e^{-i\Omega^a\hat{\sigma}_a}|
\theta ',\phi '\>
\label{6.1} \\
&\equiv & \int\limits_{}^{}\frac{d\phi d\theta}{(2\pi)^{N+M}}\<\theta ''
,\phi '',T|\theta ,\phi \>\<\theta ,\phi |\hat{\cal P}|\theta ',\phi '\>\
.
\label{6.2}
\end{eqnarray}
The unconstrained transition amplitude is given by the coherent-state path
integral
\begin{equation}
\<\theta '',\phi '',T|\theta ',\phi '\>=\int\limits_{}^{}{\cal D}\theta
{\cal D}\phi \exp \left(i\int\limits_{0}^{T}dt\left[\frac{1}{2}
\theta^i\dot{\theta}_i+\frac{1}{2}\phi ^a\dot{\phi}_a-\bar{h}_s(\theta,
\phi )\right]\right)
\label{6.6}
\end{equation}
with the boundary conditions $\theta (0)=\theta ', \ \theta(T)=\theta ''$
and $\phi (0)=\phi ',\ \phi (T)=\phi ''$; here $\theta _i=\ooo{ij}\theta
^j,\ \phi _a=\ooo{ab}\phi ^b$ and $\bar{h}_s$ is the lower symbol for the
operator $\hat{\bar{H}}_s$
\begin{equation}
\hat{\bar{H}}_s=\int\limits_{}^{}\frac{d\theta}{(2\pi )^N}
\frac{d\phi}{(2\pi)^M}
\bar{h}_s(\theta ,\phi )|\theta ,\phi \>\<\phi ,\theta |\ .
\label{6.7}
\end{equation}
Dividing the time interval $T$ into $n$ pieces $\varepsilon =T/n$ and
taking a convolution of $n$ kernels (\ref{6.2}) where
$T\rightarrow\varepsilon$, we arrive at the coherent-state 
path integral representation
of the physical transition amplitude
\begin{equation}
\<\theta'',\phi'',T|\theta',\phi'\>^{ph} =
\int {\cal D}\theta{\cal D}\phi{\cal D} C(\omega)
e^{i\tint_0^T dt \left(\frac 12\theta^i\dot{\theta}_i +
\frac{1}{2}\phi^a\dot{\phi}_a - \omega^a\sigma_a - \bar{h}_s
\right)}\ .
\label{6.8}
\end{equation}
The measure ${\cal D} C(\omega)$ for gauge variables,
being the product of the local measures $\delta_\sigma \omega^a(t)$,
provides the projection
at each moment of time. The action in the exponential in (\ref{6.8})
coincides with the classical action (\ref{5.6}) up to possible operator
ordering terms $\bar{H}_s -\bar{h}_s = O(\hbar)$. It is invariant with
respect to gauge transformations generated by $\sigma_a$
\begin{equation}
\delta\theta^i =\xi^a {\rm ad} \sigma_a\theta^i\ ,\ \ \ \
\delta\phi^a = \xi^b {\rm ad} \sigma_b \phi_a\ ,\ \ \ \
\delta\omega^a = - \dot{\xi}^a\ ,
\label{6.9}
\end{equation}
where the infinitesimal functions of time $\xi^a$ satisfy zero boundary
conditions
\begin{equation}
\xi^a(0) = \xi^a(T) =0\ ,
\label{6.10}
\end{equation}
which ensure that the boundary terms occurring upon varying $\bar{S}$
vanish.

To obtain the corresponding path integral on the physical
phase space, one usually has to integrate out all the gauge
variables $\omega$ and non-physical degrees of freedom.
Formally, it can be achieved by going over to new
canonical variables such that the abelian constraints
$\sigma_a$ become new canonical momenta. Due to the 
gauge invariance the Hamiltonian $\bar{h}_s$ is independent of
the corresponding canonical coordinates. Since the Liouville
measure constituting the formal path integral measure is
invariant under canonical transformations, the integration
over non-physical variables becomes trivial. Yet, the gauge
transformations in the new variables are translations of
canonical coordinates for the new momenta $\sigma_a$, so the
gauge average can also be done explicitly. Having done this,
one seems to obtain a path integral over the physical phase space
parametrized by a certain set of canonical variables.
Note that the canonical transformation discussed above is not
unique and is determined only up to a general canonical transformation
on the physical phase space. On the other hand, we have seen in
section 4 that quantum theory may well depend on a particular 
parametrization of the physical phase space, which is in
conflict with the formal coordinate invariance of the path
integral measure.

To explain the contradiction we observe that 
the above procedure of reducing the path integral measure
onto the physical phase space relies on the formal invariance
of the conventional Liouville measure with respect to canonical
transformation. Unfortunately, this is a wrong assumption. 
A typical example
is a Hamiltonian dynamics on a phase-space plane. 
By a canonical transformation
one can always locally turn the Hamiltonian into a free particle
Hamiltonian. So assuming a formal invariance of the path integral
measure with respect to general canonical transformations we would
arrive to a contradiction that every quantum dynamical 
system with one degree of freedom is equivalent to a free particle. 

Thus, in order to obtain a path integral on the physical phase 
space, the path integral measure in the amplitude (\ref{6.8})
should be regularized in a way that provides a true covariance
of the path integral (\ref{6.8}) with respect to general canonical
transformation.

\subsection{The Wiener measure regularized path integral}
\setcounter{equation}0

For Hamiltonian systems without constraints a regularization
of the path integral measure can be achieved by replacing
the conventional Liouville measure by a pinned Wiener
measure on continuous phase-space paths.
The Wiener measure regularized phase space path integral for a
general phase function $G(p,q)$ is then  given by \cite{2}
\begin{eqnarray}
&&\hskip-.3cm\lim_{\nu\ra\infty}{\cal M_\nu}
 \int\exp\{i\tint_0^T[p_j{\dot q}^j+{\dot G}(p,q)-h(p,q)]\,dt\}\n\\
&&\hskip1.5cm\times\exp\{-(1/2\nu)\tint_0^T[{\dot p}^2
+{\dot q}^2]\,dt\}\,{\cal D} p\,{\cal D} q\n\\
 &&\hskip.3cm=\lim_{\nu\ra\infty}(2\pi)^N e^{N\nu T/2}
 \int\exp\{i\tint_0^T[p_jdq^j+dG(p,q)-h(p,q)dt]\}\,d\mu^\nu_W(p,q)\n\\
&&\hskip.3cm=\<p'',q''|e^{-i\H T}|p',q'\>\;\ ,
\label{7.1}
\end{eqnarray}
where the last relation involves a coherent state matrix element.
Here we use the convention adopted in Section 1 that $q^j =
\theta^{2j-1}$
and $p_j = \theta^{2j}$ (cf. (\ref{coh})).
In this expression we note that $\tint p_j\,dq^j$ is a
{\it stochastic integral}, and as such we need to give it a definition.
As it stands both the It\^o (nonanticipating) rule and the Stratonovich
(midpoint) rule of definition for stochastic integrals yield the same
result (since $dp_j(t)dq^k(t)=0$ is a valid It\^o rule in these
coordinates). Under any change of canonical coordinates,
we consistently will interpret this stochastic integral
in the Stratonovich sense because it will then obey the ordinary
rules of calculus.
We also emphasize the covariance of this expression
under canonical coordinate transformations. In particular,
if ${\o p}d{\o q}=pdq+dF({\o q},q)$ characterizes a canonical
transformation from the variables $p,q$ to ${\o p},{\o q}$,
then with the Stratonovich rule the path integral becomes
\begin{eqnarray}
&&\<{\o p}'',{\o q}''|e^{-i\H T }|{\o p}',{\o q}'\>\n\\
&&\hskip.3cm=\lim_{\nu\ra\infty}(2\pi)^N e^{N\nu T/2}\int
\exp\{i\tint_0^T[{\o p}_jd{\o q}^j+d{\o G}({\o p},{\o q})-{\o h}({\o p},
{\o q})dt]\}\,d\mu^\nu_W({\o p},{\o q})\n\\
&&\hskip.3cm=\lim_{\nu\ra\infty}{\cal M_\nu}\int\exp\{i\tint_0^T[{\o p}_j
{\dot{\o q}}^j+{\dot{\o G}}({\o p},{\o q})-{\o h}({\o p},{\o q})dt]\}\n\\
&&\hskip3.2cm\times\exp\{-(1/2\nu)\tint_0^T[d\sigma({\o p},{\o q})^2/dt^2]\,
dt\}\,{\cal D}{\o p}\,{\cal D}{\o q}\,,
\label{7.2}
\end{eqnarray}
where $\o G$ incorporates both $F$ and $G$.
In this expression we have set $d\sigma({\o p},{\o q})^2=dp^2+dq^2$,
namely, the new form of the flat metric in curvilinear phase space
coordinates. We emphasize that this path integral regularization
involves Brownian motion on a flat space whatever
choice of coordinates is made. Our transformation has also made
use of the formal -- and in this case valid -- invariance of the
Liouville measure.

In order to fulfill our program of a coordinate-free path integral
representation of second-class constrained systems, 
we have to extend the Wiener measure regularization to such systems.

\subsection{The Wiener measure for second-class constraints}
\setcounter{equation}0

The regularized measure in the path integral (\ref{6.8}) is obtained
by the replacement
\begin{equation}
{\cal D}\theta{\cal D}\phi{\cal D} C(\omega)\rightarrow {\cal D}
C(\omega)d\mu_W^g(\theta,\phi,\omega)\ ,
\label{8.1}
\end{equation}
where the gauged Wiener measure $d\mu_W^g$ is to be found. The Wiener
measure regularization of the path integral should not violate
gauge invariance, therefore, we impose the condition 
\begin{equation}
\delta d\mu_W^g(\theta,\phi,\omega)=0\ ,
\label{8.2}
\end{equation}
where the operator $\delta$ is determined in (\ref{6.9}).
Since the Wiener measure provides for covariance of the path integral
(\ref{6.8}) relative to canonical transformations, we perform a
canonical transformation in (\ref{6.8}) such that $\sigma_a$ become
new canonical momenta
\begin{equation}
(\theta^i, \phi^a)\rightarrow (\pi_a=\sigma_a, y^a,\vartheta^\alpha)\ ;
\label{8.3}
\end{equation}
here $y^a$ are canonical coordinates for $\pi_a$ and $\vartheta^a$
are canonical symplectic variables on the physical phase space
(Darboux variables for the physical symplectic structure (\ref{2.5})).
In the new variables the action assumes the form
\begin{equation}
\bar{S} =\int_0^T\left( \pi_ady^a +{\textstyle\frac 12}
\vartheta^\alpha d\vartheta_\alpha -\omega^a\pi_a dt -dG -\bar{h}_sdt
\right)
\label{8.4}
\end{equation}
and for the Wiener measure we get
\begin{equation}
d\mu_W^g(\theta,\phi,\omega) = d\bar{\mu}_W^g(\pi,y,\vartheta,\omega)\ .
\label{8.5}
\end{equation}
The gauge transformations (\ref{6.9}) leave all the new variables
untouched except the $y^a$ which are shifted
\begin{equation}
\delta\pi_a =0\ ,\ \ \ \ \delta\vartheta^\alpha =0\ ,\ \ \ \
\delta y^a =-\xi^a\ .
\label{8.6}
\end{equation}
By the change of integration variables
\begin{equation}
y^a(t)\rightarrow y^a(t) -\int_t^Tdt'\omega^a(t')\ ,
\label{8.7}
\end{equation}
one can remove the dependence on $\omega^a$ of the integral (\ref{6.8})
for all intermediate moments of time $0<t<T$. However the initial
values of $y^a$ are not integration variables and therefore the average
over $\omega^a$ does not disappear without a trace. Let us  make a change
of gauge variables $\omega^a\rightarrow \dot{\omega}^a$, 
where the new variables $\omega^a$ satisfy the boundary condition
$\omega^a(T)=0$. Note that we are free to add any constant to
$\omega^a$ because it does not affect the derivative $\dot{\omega}^a$;
the boundary condition fixes this arbitrariness, providing a
one-to-one correspondence between the old and new gauge integration
variables. With this choice equation (\ref{8.7}) assumes a simple
form $y^a(t) \rightarrow y^a(t) + \omega^a(t)$.  At the
boundary $t=0$ we have
\begin{equation}
y^a(0)\rightarrow y^a(0) + \Omega^a\ ,\ \ \
\omega^a(0) =\Omega^a\ .
\label{8.8}
\end{equation}
Therefore the average measure for gauge variables is reduced to a single
average over $\Omega^a$,
\begin{equation}
{\cal D} C(\dot{\omega})\rightarrow \delta_\sigma\Omega\
\label{8.9}
\end{equation}
because $\int{\cal D} C(\dot{\omega}) = 1$.

Equation (\ref{8.2}) is easy to solve in the new canonical variables
\begin{equation}
d\bar{\mu}_W^g(\pi,y,\vartheta,\omega) =
d\tilde{\mu}_W^g
(\pi,y-\omega,\vartheta)\ ,
\label{8.10}
\end{equation}
where the gauge variables $\omega$ have been replaced by their time
derivatives as in (\ref{8.8}). 
Note that under  gauge transformation $\delta(y^a-\omega^a) =
- -\xi^a - \delta\omega^a = 0$ because $\delta\omega^a = -\xi^a$
in accordance with (\ref{6.9}) and the replacement $\omega^a
\rightarrow \dot{\omega}^a$.
Clearly, the further transformation of the $y$-integral 
to the new variables $y-\omega$ 
removes the dependence of the Wiener measure on the gauge
variables for all intermediate moments of time, i.e.,
\begin{equation}
d\tilde{\mu}_W^g
(\pi,y-\omega,\vartheta)
\rightarrow
d\tilde{\mu}_W^g
(\pi,y,\vartheta)
\label{8.11}
\end{equation}
in the path integral (\ref{6.8}). As a result of these two canonical
transformations the entire dependence
of the path integral measure on gauge variables is reduced to a single
average over a gauge orbit of the initial phase-space point with
some {\em  phase factor}
determined by the phase function $\int dG$ of the
canonical transformation (\ref{8.3}). That is, we have recovered
the projection formula (\ref{6.2}) where the projection operator
kernel is given by
\begin{equation}
\<\theta,\phi |\theta',\phi'\>^{ph}  =
\int \delta_\sigma\Omega \<\theta,\phi |\theta_\Omega', \phi_\Omega'\>
e^{i\bar{G}(\theta',\phi',\Omega)}\ ,
\label{8.11a}
\end{equation}
with $\theta_\Omega$ and $\phi_\Omega$ being gauge transformations
of the extended phase space variables generated by (\ref{5.10})
with $\xi =\Omega$ and $\bar{G}$ is $G(t=0)$ written in the initial
canonical variables.

The Wiener measure regularized path integral (\ref{6.6}) involved
in the projection formula (\ref{6.2}) should have the flat Wiener
measure on the extended phase space according to our
consideration in the previous section, that is,
\begin{equation}
d\tilde{\mu}_W^g (\pi,y,\vartheta) = d\mu_W^\nu(\theta,\phi)\ .
\label{8.12}
\end{equation}
Having established the relation between $d\tilde{\mu}^g_W$ 
and the flat-space Wiener measure $d\mu_W^\nu$, 
we can perform a canonical
transformation inverse to (\ref{8.3}) to restore the dependence
of the Wiener measure on the gauge variables and thereby to find
an explicit form of the desired gauged Wiener measure (\ref{8.1}).
Combining (\ref{8.5}), (\ref{8.10}) and (\ref{8.12})
we conclude that
\begin{equation}
d\mu_W^g(\theta,\phi,\omega) = d\mu_W^\nu(\theta_\omega,\phi_\omega)\ ,
\label{8.13}
\end{equation}
where
\begin{equation}
\theta^i_\omega = e^{\omega^a{\rm ad} \sigma_a}\theta^i\ ,\ \ \ \
\phi^a_\omega =e^{\omega^b{\rm ad} \sigma_b}\phi^a\
\label{8.14}
\end{equation}
with $\omega^a = \omega^a(t)$. The gauge invariance of the gauged Wiener
measure (\ref{8.13}) follows from the simple observation that
\begin{equation}
\delta\theta_\omega^i =\delta\phi_\omega^a =0
\label{8.15}
\end{equation}
under the gauge transformation (\ref{6.9}) where $\delta\omega \equiv
- -\xi$ according to the change of gauge variables $\omega \rightarrow
\dot{\omega}$.

Thus, the Wiener measure regularized path integral for second-class
constrained theories has the form
\begin{eqnarray}
\<\theta'',\phi'',T|\theta',\phi'\>^{ph} &=&
\int {\cal D} C(\omega)\int d\mu_W^\nu (\theta_\omega,\phi_\omega)
e^{i\tint_0^T dt \left(\frac 12\theta^i\dot{\theta}_i +
\frac{1}{2}\phi^a\dot{\phi}_a - \dot{\omega}^a\sigma_a - \bar{h}_s
\right)}\ ;\
\label{8.16}\\
d\mu_W^\nu
(\theta_\omega,\phi_\omega) &=& e^{(N+M)\nu T/2} {\cal D} \theta
{\cal D}\phi \exp\left(
- -\frac{1}{2\nu}\int_0^Tdt (\dot{\theta}_\omega^2 +\dot{\phi}_\omega^2)
\right)\ ,
\label{8.17}
\end{eqnarray}
where the limit 
$\nu\rightarrow \infty$ must be 
taken after calculating the path integral (\ref{8.16}).
In general, the gauged Wiener measure (\ref{8.17}) depends not only
on $\dot{\omega}$ but also on $\omega$ themselves, therefore, it is
not possible to remove the dependence of the action in (\ref{8.16})
on the time derivatives of the gauge variables by changing the gauge
variables back $\dot{\omega}\rightarrow\omega$ in the average measure
${\cal D} C(\omega)$, while maintaining 
the locality of the gauged Wiener measure (\ref{8.17}).

As an example consider the gauged Wiener measure for the two-dimensional
rotator (a generalization to the $N$-dimensional case is trivial
as remarked after Eq. (\ref{5.9})). 
The canonical transformation (\ref{8.3}) can be chosen as
\begin{eqnarray}
\pi_1&= &\sigma_1\ ,\ \ \ \ \ \ \hskip.5cm y^1 =Q\ ; \label{8.18} \\
\pi_2&= &\sigma_2\ ,\ \ \ \ \ \ \hskip.5cm 
y^2= \ln(|\mb{x}|/R)\ ; \label{8.19}\\
\vartheta^2 &=& (\mb{p},T\mb{x})\ ,\ \ \
\vartheta^1 =\tan^{-1}(x_2/x_1)\ ,\ \ \ \{\vartheta^1,
\vartheta^2\} =1\ .\label{8.20}
\end{eqnarray}
As expected from ${\rm ad} \sigma_a \bar{H}_s =0$, the canonical coordinates
$y^a$ are cyclic. The gauge transformations
\begin{equation}
y^1\rightarrow y^1 -\xi^1\ ,\ \ \ \ \
y^2\rightarrow y^2 -\xi^2\
\label{8.19a}
\end{equation}
induce gauge transformations of the initial canonical variables
(\ref{5.11}). Setting in (\ref{5.11}) $\xi =\omega$ we obtain
the gauged flat metric on the extended phase space that determines
the Wiener measure
\begin{eqnarray}
\int_0^Tdt (\dot{\theta}_\omega^2 +
\dot{\phi}_\omega^2)&= &
\int_0^T dt\left(\dot{\mb{p}}_\omega^2 +
\dot{\mb{x}}_\omega^2 + \dot{P}_\omega^2 +
\dot{Q}_\omega^2\right)
\label{8.21} \\
&= & \int_0^T dt g_{AB}(\Lambda) \dot{\Lambda}^A\dot{\Lambda}^B\ ,
\label{8.22}
\end{eqnarray}
where $\Lambda^A$ denotes the set of all canonical and gauge variables
$(\theta,\phi,\omega) = (\mb{p},\mb{x},P,Q,\omega)$; 
for the $N$-dimensional rotator, $\mb{p}$ and $\mb{x}$ are
$N$-dimensional vectors in (\ref{5.11}) and (\ref{8.22}).
Note that the metric
$g_{AB}$ depends generally on all the $\Lambda^A$, as well as the
components $g_{A\omega}$ and $g_{\omega\omega}$ do not vanish. Thus,
the Wiener measure depends on gauge variables and their time derivatives.

Expression (\ref{8.22}) holds for general second-class constrained
systems. Its geometrical meaning is transparent. The metric $g_{AB}$
is, by construction, degenerate along the directions traversed by gauge
transformations of the $\Lambda$. Hence the gauged Wiener measure
describes a Brownian motion (with diffusion
constant that tends to infinity) 
in the directions transverse to the gauge orbits, while
the average over the gauge variables with the measure ${\cal D} C(\omega)$
regularizes the path integral along the gauge orbits. 
An explicit construction of this measure is
discussed in section 9.

An unusual feature of the integral (\ref{8.16}) is the appearance
of the time derivatives of the gauge variables in the classical
action. This was the price we paid for locality of the Wiener
measure. One should realize that this is not always the case
for the Wiener measure in gauge theories. 
If the canonical transformations 
generated by first-class constraints were linear and preserving a bilinear
positive form on the extended phase space, then the associated Wiener
measure (\ref{8.17})  would have had no dependence on $\omega$ but
on $\dot{\omega}$ only. The latter occurs for Yang-Mills type
gauge theories \cite{klsh}. In this case the dependence on
$\dot{\omega}$ can be removed by a simple change of variables
$\dot{\omega}\rightarrow \omega$ in the gauge average integral
without violating the locality of the Wiener measure.

At first sight, the presence of the time derivatives
of the gauge variables in the classical action seems to allow
for non-physical motion with $\sigma = const \neq 0$ (a variation
of the action relative to $\omega$ leads to the equation of
motion $\dot{\sigma} =0$ rather than just $\sigma =0$). One
has however to bear in mind that equation (\ref{8.16}) describes a
quantum motion whose gauge invariance is ensured by an appropriate
average over the gauge variables. As long as the measure 
${\cal D} C(\omega)$
provides at least one gauge group average in the time interval
$0\leq t\leq T$, contributions of states with $\sigma =const \neq 0$
are projected out from the transition amplitude in full accordance
with the projection formula (\ref{6.2}).

\subsection{The average measure for gauge variables}
\setcounter{equation}0

The spectrum of the first-class constraint operators that usually
occur upon the abelian conversion of second-class constraint operators
is continuous. Therefore the average measure $\delta_\sigma\Omega$
in the projection operator (\ref{projector}) should be regularized to
provide the normalization condition (\ref{6.3}). Since the converted
constraints  are abelian, the projection operator (\ref{projector}) is 
the product of the projection operators for each independent abelian 
generator $\hat{\sigma}_a$. The latter allows us to treat the measure
$\delta_\sigma\Omega$ as the product of normalized measures for each
independent gauge variables $\Omega^a$. So we can drop the index $a$
and consider the measure only for one generator $\hat{\sigma}$.

The gauge transformations are translations of the gauge parameter
$\Omega$. Hence any regularization (a cut-off) 
of the translation invariant
measure $d\Omega$ would break the translation invariance and
therefore an explicit gauge invariance of the
path integral (\ref{8.16}). In this sense, 
the regularization would lead to a ``gauge-fixing'' term in the
effective action in the integrand in (\ref{8.16}). The gauge 
invariance of the amplitude (\ref{8.16}) is guaranteed as long as
the regularized measure for gauge variables provides at least one
projection onto the physical subspace in the time interval $t\in [0,T]$.

Consider the regularized measure of the following form
\begin{equation}
\delta_\sigma\Omega = \sqrt{\frac{m}{2\pi}}\ e^{-\frac m2 \Omega^2}
d\Omega\ ,\ \ \ \ \ m\rightarrow 0\ .
\label{9.1}
\end{equation}
Here $m$ is the regularization parameter. Clearly, the measure 
(\ref{9.1}) is normalized to unity.
Let $|\sigma\>$ be an eigenvector of the generator $\hat{\sigma}$.
Applying the projector (\ref{projector}) to it we find
\begin{eqnarray}
\hat{\cal P}|\sigma\> &=& \int \delta_\sigma \Omega 
e^{-i\Omega\hat{\sigma}} |\sigma\>\ \nonumber\\
&=&\sqrt{\frac{m}{2\pi}} \int\limits_{-\infty}^\infty d\Omega\  
e^{-\frac m2 \Omega^2 - i\Omega \sigma}|\sigma\>\nonumber\\
&=& e^{-\frac{\sigma^2}{2m}} |\sigma\>\ .
\label{9.2} 
\end{eqnarray}
Taking the limit $m\rightarrow 0$ in (\ref{9.2}) we see that for 
a hermitian operator $\hat{\sigma}$, the operator $\hat{\cal P}$
annihilates all eigenvectors of the gauge generator $\hat{\sigma}$
unless $\sigma =0$. In the latter case $\hat{\cal P}$ acts as the
unit operator, that is, it is the projector on the physical subspace.

Adopting the above regularization of the gauge average measure,
we replace $T$ in the projection formula (\ref{6.1}) by an
infinitesimal time interval $\varepsilon=T/n$ 
and construct a convolution
on $n$ infinitesimal propagators (\ref{6.1}). The result has
the form (\ref{6.8}) where the gauge variable measure is
\begin{equation}
\D C(\omega) = \prod\limits_{j=0}^{n-1} 
\sqrt{\frac{m}{2\pi}}
e^{-\frac m2 \omega_j^2} d\omega_j =
{\cal N} e^{-\int_0^T dt (m\omega^2/2)} \prod\limits_t
\sqrt{m}\ d\omega(t)\ , 
\label{9.3}
\end{equation}
where $\omega(0) = \Omega$ (to match the notations in (\ref{6.1})).
To take the continuum limit we have rescaled the gauge variables
$\omega_j \rightarrow \sqrt{\varepsilon}\omega_j$ with
$\varepsilon$ being the time slicing so that $\omega_j =
\omega(t_j)$ and $\omega(t_{j+1}) = \omega(t_j +
\varepsilon)$.

To make the gauged Wiener measure a local functional of
gauge variables, it was proposed in section 8 to change
the integration variables $\omega(t)\rightarrow \dot{\omega}(t)$.
In the time-slice approximation of the path integral the change
of gauge variables assumes the form
\begin{equation}
\omega_j \rightarrow (\omega_{j+1} - \omega_j)/\varepsilon\ ,
\ \ \ \ j=0,1, ..., n-1\ .
\label{9.4}
\end{equation}
The ``extra'' new variable $\omega_n =\omega(T)$ is
fixed by the boundary condition $\omega(T) = 0$ as suggested 
in Section 8. In the new variables the measure for gauge variables
turns into a flat-space Wiener measure for continuous
paths pinned at one point
\begin{equation}
\D C_m(\dot{\omega}) = {\cal N} \exp\left(-\frac m2 \int_0^T
dt\ \dot{\omega}^2\right) \prod_t \sqrt{m}\ d\omega(t)\ ,
\ \ \ \ \ \omega(T)=0\ ,
\label{9.5}
\end{equation}
where the index $m$ stands to emphasize the dependence of the
measure on the regularization parameter $m$.
With this choice of the measure for gauge variables, we
arrive at our coordinate-free and mathematically well-defined
formulation for the path integral representation of the second
class constrained systems.

To conclude the discussion, we note that by construction the
limits $m\rightarrow 0$ and $\nu \rightarrow \infty$ commute
in the integral (\ref{8.16}).
The latter follows from the projection formula (\ref{8.11a})
to which the path integral can be transformed by a change
of variables as has been shown in Section 8. 
The amlitude $\<\theta,\phi|\theta'_\Omega,
\phi_\Omega'\>$ can be calculated at a finite $\nu$. For any
fixed $\Omega$ it is a regular function of $\nu$ in the vicinity
of $\nu = \infty$. So, taking its gauge invariant part either
before the limit $\nu \rightarrow \infty$ or after it  
would yield the same result.
It is convenient then to make a particular choice of the
parameter $m$ to simplify the path integral measure form. 
Namely, we set
\begin{equation}
m = 1/\nu \ ,
\label{9.6}
\end{equation}
so that the path integral measure would depend only on one
parameter to be taken to infinity after performing the sum over
paths. Thus, the gauged Wiener measure assumes the following
(unified) form
\begin{equation}
d\mu_W^\nu(\theta_\omega,\phi_\omega)\D C_{1/\nu}(\dot{\omega})
= e^{(N+M)\nu T/2}e^{-\frac{1}{2\nu}\int_0^T dt
\left(\dot{\theta}_\omega^2 + 
\dot{\phi}_\omega^2 + \dot{\omega}^2\right)} \D\theta
\D\phi\D \omega\ .
\label{9.7}
\end{equation}
As has been shown in Section 8, by a suitable change of variables one
can always remove the dependence of the integrand in (\ref{8.16})
on the gauge variables $\omega^a$ for all moments of time
except $t=0$. Then the integral over $\omega^a$ yields
the kernel of the imaginary time transition amplitude for
a free motion in the M-dimensional Euclidean 
space of gauge parameters where 
$\omega^a(T) = 0$ and $\omega^a(0)= \Omega^a$. The integration
over the initial values $\Omega^a$ weighted with this
kernel is precisely the gauge average (\ref{6.1}) 
regularized as prescribed by  (\ref{9.2}). The measure (\ref{9.7})
describes a Brownian motion
with an arbitrarily large diffusion constant
on the unified space $(\theta,\phi,\omega)$, and, in this sense,
all degrees of freedom, physical and gauge ones, are treated 
on equal footing in the Wiener measure (\ref{9.7}). 

We note that Ashworth has also studied Wiener measure regularizations for 
systems with first class constraints \cite{ash}.

\end{document}